\documentclass{article}
\usepackage{graphicx}
\usepackage{amsmath}

\begin{document}

\title{Bogoliubov Approach to Superfluidity of Atoms in an Optical Lattice}
\author{$^{3}$Ana Maria Rey, $^{1}$Keith Burnett, $^{1}$Robert Roth, $^{2}$Mark
Edwards, $^{3}$ Carl J. Williams \and $^{3}$Charles W. Clark \\
%EndAName
$^{1}$Department of Physics,Clarendon Laboratory\\
Parks Road, Oxford OX1 3PU, United Kingdom\\
$^{2}$Department of Physics, Georgia Southern University\\
Statesboro, GA 30460-8031, USA\\
$^{3}$ Physics Laboratory\\
National Institute of Standards and Technology\\
Technology Administration\\
U.S. Department of Commerce\\
Gaithersburg, MD 20899-8410, USA}
\maketitle

\begin{abstract}
We use the Bogoliubov theory of atoms in an optical lattice to study the
approach to the Mott-insulator transition. We derive an explicit expression
for the superfluid density based on the rigidity of the system under phase
variations. This enables us to explore the connection between the quantum
depletion of the condensate and the quasi-momentum distribution on the one
hand and the superfluid fraction on the other. The approach to the insulator
phase may be characterized through the filling of the band by quantum
depletion, which should be directly observable via the matter wave
interference patterns. We complement these findings by self-consistent
Hartree-Fock-Bogoliubov-Popov calculations for one-dimensional lattices
including the effects of a parabolic trapping potential.
\end{abstract}

%%%%%%%%%%%%%%%%%%%%%%%%%%%%%%%%%%%%%%%%%%%%%%%%%%%%%%%%%%%%%%%%%%

\section{Introduction}

Spectacular progress has been made in experimental studies of atoms loaded
into an optical lattice in the region of the Mott superfluid insulator
quantum phase transition \cite{Kasevich, BlochMunich}. In this article, we
shall discuss the superfluid density and the quasi-momentum distribution,
which is directly related to the matter-wave interference patterns that can
be observed in such experiments. To do this we use the Bogoliubov method
\cite{Bogoliubov}, as developed for use in optical lattices \cite{VanOosten}%
. In a previous paper \cite{BurnettCorrelations} we used this method to
produce results for squeezing that are consistent with those of other
approaches previously reported in the literature \cite
{Java1,Java2,Shotter,Baym}. In this paper we shall show how it can be used
to predict the decrease in the superfluid fraction and the corresponding
variations in the matter wave interference fringes that should be directly
observable in future experiments. This extends our previous studies based on
exact calculation for small one-dimensional systems \cite{RBSuperfluid} into
the experimentally relevant regime of lattice sizes and particle numbers.

We first introduce the Bose-Hubbard Hamiltonian for atoms in an optical
lattice \cite{Jaksch}. We then describe briefly our method for determining
the superfluid fraction based on the rigidity of the system under a twist of
the condensate phase \cite{FisherBarberJasnow}. Using a perturbative
formulation analogous to the Drude weight \cite{Drude} the Bogoliubov
approximation gives us a particularly direct way of finding this quantity.
It also gives a simple picture of how superfluidity is suppressed by quantum
depletion of the condensate. We shall compare the results for various
quantities calculated using the Bogoliubov approximation with exact
numerical calculations for the case of modest numbers of atoms and lattice
sites \cite{RBSuperfluid}.

%%%%%%%%%%%%%%%%%%%%%%%%%%%%%%%%%%%%%%%%%%%%%%%%%%%%%%%%%%%%%%%%%%

\section{The Bose-Hubbard Model and Superfluidity}

The Bose-Hubbard Hamiltonian for atoms in an one-dimensional optical lattice
with $I$ sites has the form \cite{Jaksch}: %%
\begin{equation}
\hat{H} = \sum_{i=1}^{I}\hat{n}_{i}\epsilon_{i} - J \sum_{i=1}^{I}(\hat{a}%
_{i+1}^{\dagger}\hat{a}_{i} + \hat{a}_{i}^{\dagger}\hat{a}_{i+1}) + \frac{V}{%
2} \sum_{i=1}^{I} \hat{n}_{i}(\hat{n}_{i}-1) \;.  \label{hamil}
\end{equation}
Here $J$ represents the coupling between adjacent lattice sites
due to tunneling and $V$ is the strength of repulsion between
atoms on the same site. The non-interacting energy of the atoms on
each site, $\epsilon_{i}$, will have some variation that is
typically smooth on the scale of the condensate. We shall consider
below both, the case where this is a constant, as well as the
extension to the case where it varies in a trapped condensate.
This Bose-Hubbard Hamiltonian should be an appropriate model when
the loading process produces atoms in the lowest vibrational state
of each well, with a chemical potential smaller than the distance
to the first vibrationally excited state. This is known to be
possible from the results of recent experiments
\cite{Kasevich,BlochMunich,NISTLattice}.

%In the following we want to use the Bogoliubov approach to investigate the
%superfluid properties of these lattice systems.  One possible microscopic
%definition of superfluidity is based on the response of the system to an
%imposed variation of the phase of the wave function \cite{Legett}. Any
%spatial variation of the phase is associated with a velocity field
%proportional to the gradient of the phase. This irrotational velocity
%field is identified with the non-dissipative superfluid flow
%\cite{LifshitzandPitaevskii}. Due to the kinetic energy of this superflow
%the energy $E_{\theta}$ of the system with an imposed phase twist is
%larger than the energy $E_{0}$ of the system without phase variation. This
%energy difference is a direct measure for the superfluid fraction
%\cite{Krauth,FisherBarberJasnow,RBSuperfluid}
%%
%\begin{equation}
%f_{\mathrm{s}} = \frac{1}{N}\frac{E_{\theta}-E_{0}}{J (\Delta\theta)^{2}} \;.
%\end{equation}
%%
%For simplicity we assume a linear phase variation across the lattice with
%a total twist angle $\theta$ and $\Delta\theta=\theta/I$. This relation
%reveals that superfluidity is connected to the rigidity of the
%lattice system under an imposed phase twist.

The concept of superfluidity is closely related to the existence of a
condensate in the interacting many--body system. Formally, the one--body
density matrix $\rho ^{(1)}\left( \vec{x},\vec{x}^{\prime }\right) $ has to
have exactly one macroscopic eigenvalue which defines the number of
particles in the condensate; the corresponding eigenvector describes the
condensate wave function $\phi _{0}\left( \vec{x}\right) =e^{i\Theta (\vec{x}%
)}\left| \phi _{0}\left( \vec{x}\right) \right| $. A spatially varying
condensate phase, $\Theta \left( \vec{x}\right) $, is associated with a
velocity field for the condensate by
\begin{equation}
\vec{v}_{0}\left( \vec{x}\right) =\frac{\hbar }{m}\vec{\nabla}\Theta \left(
\vec{x}\right) .  \label{ediff}
\end{equation}
This irrotational velocity field is identified with the velocity of the
\emph{superfluid flow}, $\vec{v}_{s}\left( \vec{x}\right) \equiv \vec{v}%
_{0}\left( \vec{x}\right) $
(\cite{Legett},\cite{LifshitzandPitaevskii}) and enables us to
derive an expression for the superfluid fraction, $f_{s}$.
Consider a system with a finite linear dimension, $L$, in the
$\vec{e}_{1}$--direction and a ground--state energy, $E_{0}$,
calculated
with periodic boundary conditions. Now we impose a linear phase variation, $%
\Theta \left( \vec{x}\right) =\theta x_{1}/L$ with a total twist angle $%
\theta $ over the length of the system in the $\vec{e}_{1}$--direction. The
resulting ground--state energy, $E_{\theta }$ will depend on the phase
twist. For very small twist angles, $\theta \ll \pi $, the energy
difference, $E_{\theta }-E_{0}$, can be attributed to the kinetic energy, $%
T_{s}$, of the superflow generated by the phase gradient. Thus,
\begin{equation}
E_{\theta }-E_{0}=T_{s}=\frac{1}{2}mNf_{s}\vec{v}_{s}^{2},
\end{equation}
where $m$ is the mass of a single particle and $N$ is the total
number of particles so that $mNf_{s}$ is the total mass of the
superfluid component. Replacing the superfluid velocity,
$\vec{v}_{s}$ with the phase gradient according to Eq.\
(\ref{ediff}) leads to a fundamental relation for the superfluid
fraction
\begin{equation}
f_{s}=\frac{2m}{\hbar ^{2}}\frac{L^{2}}{N}\frac{E_{\theta }-E_{0}}{\theta
^{2}}=\frac{1}{N}\frac{E_{\theta }-E_{0}}{J\left( \Delta \theta \right) ^{2}}%
,
\end{equation}
where the second equality applies to a lattice system on which a linear
phase variation has been imposed. Here the distance between sites is $a$,
the phase variation over this distance is $\Delta \theta $, and the number
of sites is $I$. In this case, $J\equiv \hbar ^{2}/(2ma^{2})$.

Technically the phase variation can be imposed through so-called twisted
boundary conditions \cite{FisherBarberJasnow}. In the context of the
discrete Bose-Hubbard model it is, however, more convenient to map the phase
variation by means of a unitary transformation onto the Hamiltonian. The
resulting ``twisted'' Hamiltonian %%
\begin{equation}
\hat{H}_{\theta} = \sum_{i=1}^{I} \hat{n}_{i} \epsilon_{i} - J
\sum_{i=1}^{I} (e^{-\mathrm{i}\Delta \theta}
\hat{a}_{i+1}^{\dagger }\hat{a} _{i} + e^{\mathrm{i}\Delta \theta}
\hat{a}_{i}^{\dagger} \hat{a}_{i+1}) + \frac{V}{2} \sum_{i=1}^{I}
\hat{n}_{i}(\hat{n}_{i}-1) \, \label{hamil_twist}
\end{equation}
exhibits additional phase factors $e^{\pm\mathrm{i}\Delta\theta}$ --- the
so-called Peierls phase factors --- in the hopping term \cite
{Poilblanc,ShastrySutherland}. These phase factors show that the twist is
equivalent to the imposition of an acceleration on the lattice for a finite
time. It is interesting to note that the present experiments enable us to
make a specific connection between the formal and operational aspects of the
system.

%
% We want to use the Bogoliubov approach to calculate the superfluid fraction
% based on the rigidity of the lattice system under an imposed phase twist
% \cite{FisherBarberJasnow,RBSuperfluid}. The superfluid fraction can be found
% from the energy of the system where the Hamiltonian is modified by a phase
% change $\Delta\theta$ between adjacent sites, as follows: %%
% \begin{equation}
% \hat{H}_{\theta} = \sum_{i=1}^{I} \hat{n}_{i} \epsilon_{i} - J
% \sum_{i=1}^{I} (e^{-\mathrm{i}\Delta \theta} \hat{a}_{i+1}^{\dagger }\hat{a}%
% _{i} + e^{\mathrm{i}\Delta \theta} \hat{a}_{i}^{\dagger} \hat{a}_{i+1}) +
% \frac{V}{2} \sum_{i=1}^{I} \hat{n}_{i}(\hat{n}_{i}-1) \;.
% \label{hamil_twist}
% \end{equation}
% %%
%
% In the formal theory of defining the superfluid fraction this is equivalent
% to finding the ground state of the system with twisted boundary conditions.
% In the operational sense, this twist is equivalent to the imposition of an
% acceleration on the lattice for a finite time. It is interesting to note
% that the present experiments enable us to make a specific connection between
% the formal and operational aspects of the system. The superfluid fraction is
% given by the following expression \cite{Krauth,RBSuperfluid}: %%
% \begin{equation}
% f_{\mathrm{s}} = \frac{1}{N}\frac{E_{\theta}-E_{0}}{J (\Delta\theta)^{2}} \;.
% \end{equation}
% %%
% Here, the $E_{\theta}, E_{0}$ indicate respectively the ground state energy
% of the ``twisted'' or original Hamiltonian. We associate the change in
% energy solely with the kinetic energy of the superflow generated by the
% phase twist.

We calculate the change in energy $E_{\theta}-E_{0}$ under the assumption
that the phase change $\Delta \theta $ is small so that we can write: %%
\begin{equation}
e^{-\mathrm{i}\Delta \theta }\simeq 1-\mathrm{i}\Delta \theta -\frac{1}{2}%
(\Delta \theta )^{2}\;.
\end{equation}
Using this expansion the twisted Hamiltonian (\ref{hamil_twist}) takes the
following form: %%
\begin{equation}
\hat{H}_{\theta }\simeq \hat{H}_{0}+\Delta \theta \hat{J}-\frac{1}{2}(\Delta
\theta )^{2}\hat{T}=\hat{H}_{0}+\hat{H}_{\text{pert}}\;,
\end{equation}
where we retain terms up to second order in $\Delta \theta $. The current
operator $\hat{J}$ (N.B. that the physical current is given by this
expression multiplied by $\frac{1}{\hbar }$) and the hopping operator $\hat{T%
}$ are given by: %%
\begin{eqnarray}
\hat{J} &=&\mathrm{i}J\sum_{i=1}^{I}(\hat{a}_{i+1}^{\dagger }\hat{a}_{i}-%
\hat{a}_{i}^{\dagger }\hat{a}_{i+1}) \\
\hat{T} &=&-J\sum_{i=1}^{I}(\hat{a}_{i+1}^{\dagger }\hat{a}_{i}+\hat{a}%
_{i}^{\dagger }\hat{a}_{i+1})\;.
\end{eqnarray}
The change in the energy $E_{\theta }-E_{0}$ due to the imposed phase twist
can now be evaluated in second order perturbation theory %%
\begin{equation}
E_{\theta }-E_{0}=\Delta E^{(1)}+\Delta E^{(2)}\;.
\end{equation}
The first order contribution to the energy change is proportional to the
expectation value of the hopping operator %%
\begin{equation}
\Delta E^{(1)}=\langle \Psi _{0}|\hat{H}_{\text{pert}}|\Psi _{0}\rangle =-%
\frac{1}{2}(\Delta \theta )^{2}\langle \Psi _{0}|\hat{T}|\Psi _{0}\rangle \;.
\end{equation}
Here $|\Psi _{0}\rangle $ is the ground state of the original Bose-Hubbard
Hamiltonian (\ref{hamil}). The second order term is related to the matrix
elements of the current operator involving the excited states $|\Psi _{\nu
}\rangle $ ($\nu =1,2,...$) of the original Hamiltonian %%
\begin{equation}
\Delta E^{(2)}=-\sum_{\nu \neq 0}\frac{|\langle \Psi _{\nu }|\hat{H}_{\text{%
pert}}|\Psi _{0}\rangle |^{2}}{E_{\nu }-E_{0}}=-(\Delta \theta
)^{2}\sum_{\nu \neq 0}\frac{|\langle \Psi _{\nu }|\hat{J}|\Psi _{0}\rangle
|^{2}}{E_{\nu }-E_{0}}\;.
\end{equation}
Thus we obtain for the energy change up to second order in $\Delta \theta $
\begin{eqnarray}
E_{\theta }-E_{0}&=&(\Delta \theta )^{2}\Big(-\frac{1}{2}\langle \Psi _{0}|%
\hat{T}|\Psi _{0}\rangle -\sum_{\nu \neq 0}\frac{|\langle \Psi _{\nu }|\hat{J%
}|\Psi _{0}\rangle |^{2}}{E_{\nu }-E_{0}}\Big)=I(\Delta \theta )^{2}D ,
\notag \\
D&\equiv&\frac{1}{I}\Big(-\frac{1}{2}\langle \Psi _{0}|\hat{T}|\Psi
_{0}\rangle -\sum_{\nu \neq 0}\frac{|\langle \Psi _{\nu }|\hat{J}|\Psi
_{0}\rangle |^{2}}{E_{\nu }-E_{0}}\Big).
\end{eqnarray}
The quantity $D$, defined above, is formally equivalent to the
Drude weight used to specify the DC conductivity of charged
fermionic systems \cite{Drude}. The superfluid fraction is then
given by the contribution of both the first and second order term:
\begin{eqnarray}
f_{\mathrm{s}}&=& f^{(1)}_{\mathrm{s}}-f^{(2)}_{\mathrm{s}}\label{sffraction};\notag\\
f^{(1)}_{\mathrm{s}}&\equiv&-\frac{1}{2NJ}\Big(\langle \Psi _{0}|%
\hat{T}|\Psi _{0}\rangle\Big),\\
f^{(2)}_{\mathrm{s}}&\equiv&\frac{1}{NJ}\Big(\sum_{\nu \neq 0}\frac{|\langle \Psi _{\nu }|\hat{J%
}|\Psi _{0}\rangle |^{2}}{E_{\nu }-E_{0}}\Big).\notag
\end{eqnarray}
Here $N$ is the number of atoms in the lattice. In general both, the first
and the second order term contribute. For a translationally invariant
lattice the second term vanishes (as is going to be shown latter) in the
Bogoliubov limit that we shall use in this study. However, in exact
calculations and in the Bogoliubov approximation for an inhomogeneous
lattice the second order term plays a role.

We can further understand this approach to the superfluid density by
calculating the flow that is produced by the application of the phase twist.
To do this we work out the expectation value of the current operator
expressed in terms of the twisted variables: %%
\begin{equation}
\hat{J}_{\theta} = \mathrm{i} J \sum_{i=1}^{I} (e^{-\mathrm{i} \Delta\theta}
\hat{a}_{i+1}^{\dagger}\hat{a}_{i} -e^{\mathrm{i} \Delta\theta} \hat{a}%
_{i}^{\dagger}\hat{a}_{i+1}) \;.
\end{equation}
We expand this to find the lowest order contributions, i.e.: %%
\begin{equation}
\hat{J}_{\theta} \simeq \hat{J} + J \Delta\theta \sum_{i=1}^{I} (\hat{a}%
_{i+1}^{\dagger} \hat{a}_{i} + \hat{a}_{i}^{\dagger }\hat{a}_{i+1}) = \hat{J}
- \hat{T} \Delta\theta \;.
\end{equation}
We use first order perturbation theory on the wave function to obtain the
following expression: %%
\begin{eqnarray}
\langle \Psi(\Delta\theta)| \hat{J}_{\theta} |\Psi(\Delta\theta)\rangle
&=&2\Delta\theta \Big(-\frac{1}{2} \langle \Psi_{0}| \hat{T}
|\Psi_{0}\rangle - \sum_{\nu\ne0} \frac{| \langle\Psi_{\nu}| \hat{J}
|\Psi_{0}\rangle |^{2}}{ E_{\nu}-E_{0}} \Big) \\
&=&2 N J f_{\mathrm{s}}\Delta\theta .  \label{flux}
\end{eqnarray}
If we note that the kinetic energy for a small quasi-momentum $q$ on a
lattice is given by $Jq^{2}a^{2}$, we can define the effective mass as $%
m^{*}=\frac{\hbar^{2}}{2J a^{2}}$. Here the quasi-momenta are given by $q =
\frac{2\pi}{Ia}\, j$ with $j=1,...,(I-1)$ and lattice spacing $a$. Thus, the
physical current, Eq. (\ref{flux}) multiplied by $\frac{1}{\hbar}$, can be
expressed as: %%
\begin{equation}
\langle \Psi(\Delta\theta)| \hat{J}_{\theta} |\Psi(\Delta\theta)\rangle = N
f_{\mathrm{s}} \Delta\theta \frac{\hbar}{m^{*}a^{2}} \;.
\end{equation}
This is the total flux and we need to divide $I$ to get the flux density,
i.e. %%
\begin{eqnarray}
\frac{1}{I}\left\langle \Psi (\Delta \theta )\right| \hat{J}_{\theta}\left|
\Psi (\Delta \theta )\right\rangle &=& \left(\frac{\hbar\Delta\theta}{m^{*}a}%
\right) \left(\frac{Nf_{\mathrm{s}}}{a I}\right)  \notag \\
&=&v_{\text{s}}n_{\text{s}} \;.
\end{eqnarray}
So we see that the Drude formulation of the superfluid fraction (\ref
{sffraction}) gives an intuitively satisfying expression for the amount of
flowing superfluid.

%%%%%%%%%%%%%%%%%%%%%%%%%%%%%%%%%%%%%%%%%%%%%%%%%%%%%%%%%%%%%%%%%%%%%%%%%

\section{The Bogoliubov Approximation to the Bose-Hubbard Hamiltonian}

We use the Bogoliubov approximation for the Bose-Hubbard model in the limit
that quantum fluctuations, or equivalently depletion of the condensate, is
not too great. In the limit that the quantum depletion can be completely
ignored, we can replace the creation and annihilation operators, $\hat{a}%
^{\dag}_i$ and $\hat{a}_i$, on each site with a $c$-number, $z_{i}$. This
leads to a set of coupled non-linear Schr\"odinger, i.e. Gross-Pitaevskii
(GP), equations for these amplitudes \cite{BHMeanField}: %%
\begin{equation}
\mathrm{i} \hbar\, \partial_{t} z_{i} = -J (z_{i+1}+z_{i-1}) + Vz_{i}
z_{i}^{\ast} z_{i} \;.
\end{equation}
This equation can be used to study the properties of the condensate loaded
into the lattice when the tunneling kinetic energy is large enough compared
to the interaction energy though small enough for the one-band Bose-Hubbard
model to be valid. We then include the quantum fluctuations in our
description of the system using the Bogoliubov approximation, where we
suppose that we can write the full annihilation operator in terms of the $c$%
-number part and a fluctuation operator thus: %%
\begin{equation}
\hat{a}_{i} = (z_{i}+\hat{\delta}_{i}) e^{-\mathrm{i}\frac{\mu t}{\hbar}} \;.
\end{equation}
This form will be useful when we are looking at the properties of a
time-independent or adiabatic ground state. In using this method we are
assuming that the fluctuation part is small. The Bogoliubov method gives us
expressions for the averages of the squares of the fluctuation operator and
allows us to determine whether this assumption is valid. We shall examine
its validity by comparing the results for various physical quantities with
exact numerical calculations based on the Bose-Hubbard Hamiltonian.

%%%%%%%%%%%%%%%%%%%%%%%%%%%%%%%%%%%%%%%%%%%%%%%%%%%%%%%%%%%%%%%%%%%%%%%%%

\subsection{Bogoliubov theory for the translationally invariant lattice}

The ground state solution for the translationally invariant lattice gives
the eigenvalue: %%
\begin{equation}
\mu = n_{0} V - 2J,
\end{equation}
where %%
\begin{equation}
n_{0}=N/I
\end{equation}
is the mean number of atoms on each site of the lattice. We take $N$ to be
the total number of atoms and $I$ to be the number of sites in the
one-dimensional lattice.

The Bogoliubov equations for the lattice have the following form: %%
\begin{equation}
\mathrm{i}\hbar\, \partial_{t}\hat{\delta}_{i} = (2n_{0}V-\mu) \hat{\delta}%
_{i} - J(\hat{\delta}_{i+1} + \hat{\delta}_{i-1}) + n_{0}V \hat{\delta}%
_{i}^{\dagger} \;.
\end{equation}
This is solved by constructing quasi-particles for the lattice which
diagonalize the Hamiltonian \cite{VanOosten}, i.e %%
\begin{eqnarray}
\hat{\delta}_{i} &=& \frac{1}{\sqrt{I}} \sum_{q} [u^{q}\hat{\alpha}_{q}\; e^{%
\mathrm{i}(qia-\omega _{q}t)} - v^{q
\ast}\hat{\alpha}_{q}^{\dagger}\; e^{-\mathrm{i}(qia-\omega
_{q}t)}]
\label{qua1} \\
\hat{\delta}_{i}^{\dagger} &=& \frac{1}{\sqrt{I}} \sum_{q} [u^{q \ast }\hat{%
\alpha}_{q}^{\dagger}\;e^{-\mathrm{i}(qia-\omega _{q}t)} - v^{q}\hat{\alpha}_{q}\; e^{%
\mathrm{i}(qia-\omega _{q}t)}] \;,  \label{qua2}
\end{eqnarray}
where $a$ is the lattice spacing. The quasi-particle operators obey the
usual Bose commutation relations: %%
\begin{equation}
\big[ \hat{\alpha}_{q},\hat{\alpha }_{q^{\prime}}^{\dagger}\big] =
\delta_{qq^{\prime}}
\end{equation}
and have the following expectation values at some temperature $T$: %%
\begin{equation}
\langle\hat{\alpha} _{q}^{\dagger} \hat{\alpha}_{q^{\prime}}\rangle =
\delta_{qq^{\prime}} [\exp(\hbar\omega_{q}/k_{b}T)-1]^{-1}.
\end{equation}

We then find the following equations for the excitation amplitudes and
frequencies, %%
\begin{eqnarray}
\hbar \omega_{q} u^{q} &=& \Big[ n_{0}V + 4J \sin^{2} \Big(\frac{qa}{2}\Big) %
\Big] u^{q} - n_{0} V v^{q}, \\
-\hbar \omega_{q} v^{q} &=& \Big[ n_{0} V + 4J \sin^{2}\Big(\frac{qa}{2}\Big)%
\Big] v^{q} - n_{0} V u^{q}.
\end{eqnarray}
Thus, the expressions for the $u^{q}$ and $v^{q}$ yield: %%
\begin{eqnarray}
|u^{q}|^{2} &=& \frac{K(q)+n_{0}V+\hbar \omega _{q}}{2\hbar \omega_{q}} \\
|v^{q}|^{2} &=& \frac{K(q)+n_{0}V-\hbar \omega _{q}}{2\hbar \omega_{q}} \;,
\end{eqnarray}
where the phonon excitation frequencies are given by: %%
\begin{eqnarray}
\hbar \omega _{q} &=& \sqrt{K(q)[2n_{0}V+K(q)]} \\
K(q) &=& 4J \sin^{2}\Big(\frac{qa}{2}\Big) \;.  \label{kq}
\end{eqnarray}

%%%%%%%%%%%%%%%%%%%%%%%%%%%%%%%%%%%%%%%%%%%%%%%%%%%%%%%%%%%%%%%%%%%%%%

\subsection{Expressions for the number superfluid fraction in the
translationally invariant lattice}

Having obtained the expressions for the excitations we can now determine the
superfluid fraction. The quantity we need to calculate is just the first
order term of the Drude expression (\ref{sffraction}), because the second
order term vanishes in the Bogoliubov limit due to the translational
invariance of the lattice [see Eq. (\ref{HFBsf})], i.e. %%
\begin{equation}
f_{\mathrm{s}} = -\frac{1}{2NJ} \langle\Psi_{0}| \hat{T} |\Psi_{0}\rangle =%
\frac{1}{2N}\sum_{i=1}^{I} \langle\Psi_{0}| \hat{a}_{i+1}^{\dagger}\hat{a}%
_{i} + \hat{a}_{i}^{\dagger}\hat{a}_{i+1} |\Psi_{0}\rangle \;.
\end{equation}
In the Bogoliubov approximation this has the form: %%
\begin{eqnarray}
f_{\mathrm{s}} &=& \frac{1}{2N}\sum_{i=1}^{I} \langle\Psi_{0}| (\hat{\delta}%
_{i+1}^{\dagger}+z_{i+1})(\hat{\delta}_{i}+z_{i}) +(\hat{\delta}%
_{i}^{\dagger}+z_{i})(\hat{\delta}_{i+1}+z_{i+1}) |\Psi_{0}\rangle \notag \\
&=& \frac{1}{2N} \sum_{i=1}^{I} \langle\Psi_{0}| 2z_{i}^{2} + \hat{\delta}%
_{i+1}^{\dagger} \hat{\delta}_{i} + \hat{\delta}_{i}^{\dagger} \hat{\delta}%
_{i+1} |\Psi_{0}\rangle \;.
\end{eqnarray}
We can now express the fluctuation operators, Eqs. (\ref{qua1}) and (\ref
{qua2}), in terms of the quasi-particle operators that diagonalize the
quadratic Hamiltonian. This leads to the following expression for the
superfluid fraction at finite temperature: %%
\begin{eqnarray}
f_{\mathrm{s}} &=& \frac{1}{2N} \bigg[ \sum_{i=1}^{I} 2 z^{2}_{i} + \frac{1}{%
I} \Big\langle \sum_{q} [u^{q}\hat{\alpha}_{q}\; e^{\mathrm{i}q(i+1)a} -
v^{q}\hat{\alpha}_{q}^{\dagger}\; e^{-\mathrm{i}q(i+1)a}]  \notag \\
&&\qquad\times \sum_{q^{\prime}}[u^{q^{\prime}}\hat{\alpha}%
_{q^{\prime}}^{\dagger} \;e^{-\mathrm{i}q^{\prime}ia} - v^{q^{\prime}}\hat{%
\alpha}_{q^{\prime}}\; e^{\mathrm{i}q^{\prime}ia}] \Big\rangle  \notag \\
&&\qquad+ \frac{1}{I} \sum_{i=1}^{I} \Big\langle \sum_{q} [u^{q}\hat{\alpha}%
_{q}^{\dagger}\;e^{-\mathrm{i}qia} - v^{q}\hat{\alpha}_{q}\; e^{\mathrm{i}%
qia}]  \notag \\
&&\qquad\times \sum_{q^{\prime}} [u^{q^{\prime}}\hat{\alpha}_{q^{\prime}}\;
e^{\mathrm{i}q^{\prime}(i+1)a} - v^{q^{\prime}}\hat{\alpha}%
_{q^{\prime}}^{\dagger}\; e^{-\mathrm{i}q^{\prime}(i+1)a}] \Big\rangle %
\bigg] \;,
\end{eqnarray}
and we find in the zero temperature limit of a translationally
invariant lattice: %%
\begin{equation}
f_{\mathrm{s}} = \frac{I}{N} \Big[z^{2} + \frac{1}{I} \sum_{q} |v^{q}|^{2}
\cos(qa) \Big] \;.  \label{sffrac_bog}
\end{equation}
Here the summation runs over all quasi-momenta $q = \frac{2\pi}{Ia}\, j$
with $j=1,...,(I-1)$ and we have called $z$ the value of all $z_{i}$ in a
translationally invariant system. This shows that in the limit of zero
lattice spacing (while keeping $q$ finite) the superfluid fraction is unity
as we have the normalization condition: %%
\begin{equation}
Iz^{2}+\sum_{q}|v^{q}|^{2}= N \;.
\end{equation}
These expressions give a direct insight into the change of the superfluid
fraction as atoms are pushed out of the condensate due to interactions. In
Eq. (\ref{sffrac_bog}) the sum involving the Bogoliubov amplitudes $v^{q}$
characterizes the difference between the condensate fraction, which is given
by the first term, and the superfluid fraction. For weak interactions and a
small depletion, which fills only the lower quarter of the band where the $%
\cos(qa)$ term has a positive sign, the superfluid fraction is larger than
the condensate fraction. Thus the depletion of the condensate has initially
little effect on superfluidity. When the depleted population spreads into
the central part of the band, where the $\cos(qa)$ term has a negative sign,
the superfluid fraction is reduced and might even become smaller than the
condensate fraction. Finally, the population in the upper quarter of the
band again produces a positive contribution to the superflow. In a sense the
interactions are playing a role akin to Fermi exclusion ''pressure'' in the
case of electron flow in a band. This, however can lead to perfect filling
and cancellation of the flow. In the case of our Bogoliubov description we
can only see reduction of the flow, not a perfect switching off of the
superfluid. This happens in the Mott insulator state, which cannot be
described by the Bogoliubov approximation.

In the next section we outline the version of the Bogoliubov theory that
should be best suited to treating these problems, i.e self-consistent
Bogoliubov theory.

%%%%%%%%%%%%%%%%%%%%%%%%%%%%%%%%%%%%%%%%%%%%%%%%%%%%%%%%%%%%%%%%%%%%%%

\section{Self-consistent HFB-Popov theory}

In this section we explore the limits of validity of the simplest zero
temperature self-consistent Bogoliubov theory, a simplified version of the
Hartree-Fock-Bogoliubov approximation originally introduced by Popov \cite
{Popov}. The HFB-Popov theory is an extension of the standard Bogoliubov
approximation that takes into account the depletion of the condensate but
neglects the anomalous average. As discussed in the previous section, taking
into account the depletion of the condensate is important as the transition
is approached because the depleted population causes the reduction of the
superfluidity. Although the HFB-Popov aproach has the limitation that it
doesn't take into account the full effect of the medium because it neglects
the anomalous average \cite{BurnettComparisons}, it can be considered a
better theory for the elementary excitations than the full HFB due to the
fact that the theory is gapless and doesn't violate Goldstone's theorem.

A derivation of the Bogoliubov equations for the quasiparticle
amplitudes in a lattice can be found for example in Ref.
\cite{BurnettCorrelations}. These equations only take into account
terms up to second order in the fluctuations. Including third and
fourth order terms by treating them in a self-consistent mean
field approximation ( \cite{Griffin}, \cite {Morgan} ) and
neglecting anomalous average terms yields the following HFB-Popov
equations:
\begin{eqnarray}
\hbar \omega _{q}u_{i}^{q}+c^{q}z_{i} &=&(2V(|z_{i}|^{2}+\tilde{n}_{i})-\mu
+\epsilon _{i})u_{i}^{q}-J(u_{i+1}^{q}+u_{i-1}^{q})-Vz_{i}^{2}v_{i}^{q}, \\
-\hbar \omega _{q}v_{i}^{q}-c^{q}z_{i}^{\ast } &=&(2V(|z_{i}|^{2}+\tilde{n}%
_{i})-\mu +\epsilon _{i})v_{i}^{q}-J(v_{i+1}^{q}+v_{i-1}^{q})-Vz_{i}^{\ast
2}u_{i}^{q}, \\
\mu z_{i} &=&-J(z_{i+1}+z_{i-1})+\left( V(|z_{i}|^{2}+2\tilde{n}%
_{i})+\epsilon _{i}\right) z_{i}, \\
\tilde{n}_{i} &=&\sum_{q}\left| v_{i}^{q}\right| ^{2}, \\
N &=&\sum_{i=1}^{I}(|z_{i}|^{2}+\tilde{n}_{i})\;, \\
c^{q} &=&V\sum_{i}|z_{i}|^{2}(z_{i}^{\ast }u_{i}^{q}-z_{i}v_{i}^{q}).
\end{eqnarray}
Where $\epsilon _{i}$ is the energy offset at site $i$ due to an
external potential, $\left\{u_{i}^{q}, v_{i}^{q}\right\}$ and
$\omega _{q}$ are respectively the quasiparticle amplitudes and
energies, thus
\begin{equation}
\hat{\delta}_{i}=\sum_{q}u_{i}^{q}\hat{\alpha}_{q}e^{-\mathrm{i}\omega
_{q}t}-v_{i}^{q\ast }\hat{\alpha}_{q}^{\dagger }e^{\mathrm{i}\omega _{q}t}\;,
\end{equation}
$\tilde{n}_{i}$ is the depletion at site $i$, and $N$ is the total
number of particles. The parameters $c^{q}$ ensure  the
$\left\{u_{i}^{q}, v_{i}^{q}\right\}$ solutions to the above
equations with $\omega _{q}\neq 0$ to be orthogonal to the
condensate (Ref. \cite{Morgan}).

By calculating the quasiparticle amplitudes and the condensate density it is
possible to get information about most of the physical properties of the
system. For example the superfluid fraction and the on site number
fluctuations are given by:
\begin{eqnarray}
f_{\mathrm{s}} &=&f_{s}^{(1)}-f_{s}^{(2)},  \label{HFBsf} \\
f_{s}^{(1)} &=&\sum_{i=1}^{I}f_{si}^{(1)}=\frac{1}{2N}\sum_{i=1}^{I}\Big[%
(z_{i+1}z_{i}^{\ast }+z_{i+1}^{\ast }z_{i})+\sum_{q}(v_{i}^{q}v_{i+1}^{q\ast
}+v_{i}^{q\ast }v_{i+1}^{q})\Big],  \notag \\
{f}_{s}^{(2)} &=&\frac{J}{N}\left( \sum_{q,q^{\prime }}\frac{\big|%
\sum_{i}\big(u_{i+1}^{q}v_{i}^{q^{\prime }}-u_{i}^{q}v_{i+1}^{q^{\prime }}%
\big)\big|^{2}}{\hbar \omega _{q}+\hbar \omega _{q^{\prime }}}+\delta
_{qq^{\prime }}\frac{\big|\sum_{i}(u_{i+1}^{q}v_{i}^{q}-u_{i}^{q}v_{i+1}^{q})%
\big|^{2}}{2\hbar \omega _{q}}\right),  \notag
\end{eqnarray}
\begin{equation}
\Delta n_{i}^{2}=|z_{i}|^{2}\sum_{q}|u_{i}^{q}-v_{i}^{q\ast
}|^{2}. \label{numflu}
\end{equation}
From the complete expression of the superfluid fraction, it can be seen
explicitly how due to the translational invariance, the second order term
vanishes in the homogeneous system.

\subsection{Translationally Invariant lattice}

For the translational invariant lattice we use the quasiparticle
transformation given by Eqs. (\ref{qua1}) and (\ref{qua2}). Under this
transformation the self consistent equations can be written, generalizing
the previous version, as:
\begin{eqnarray}
\mu &=&\left( |z|^{2}+\frac{2}{I}\sum_{q}|v^{q}|^{2}\right) V-2J, \\
|u^{q}|^{2} &=&\frac{K(q)+|z|^{2}V+\hbar \omega _{q}}{2\hbar \omega _{q}},
\label{ushfb} \\
|v^{q}|^{2} &=&\frac{K(q)+|z|^{2}V-\hbar \omega _{q}}{2\hbar \omega _{q}},
\label{eqv} \\
N &=&I|z|^{2}+\sum_{q}|v^{q}|^{2} .  \label{eqz}
\end{eqnarray}
Here the phonon excitation spectrum is given by:
\begin{equation}
\hbar \omega _{q}=\sqrt{K(q)[2|z|^{2}V+K(q)]},
\end{equation}
and $K(q)$ is given by Eq. (\ref{kq}). Again we omit the subscript
in the amplitudes $z_{i}$ because they have the same value at all
lattice sites. Notice also that due to translational invariance,
the $c^{q}$ coefficients vanish.

In the homogeneous system, the form of the HFB-Popov equations for the
quasiparticle amplitudes and energies is very close to the standard
Bogoliubov form. We do, however, have to replace $n_{0}=N/I$ by the
condensate amplitude $|z|^{2}$ which must take into account the depletion of
the condensate. We solve for the condensate amplitude as a function of the
external parameters $J,V,N$ and $I$ by inserting Eq. (\ref{eqv}) in Eq. (\ref
{eqz}). Once $|z|^{2}$ is determined, we use it to calculate the other
expressions.

\begin{figure}[tbh]
{\includegraphics[width=7.2 in]{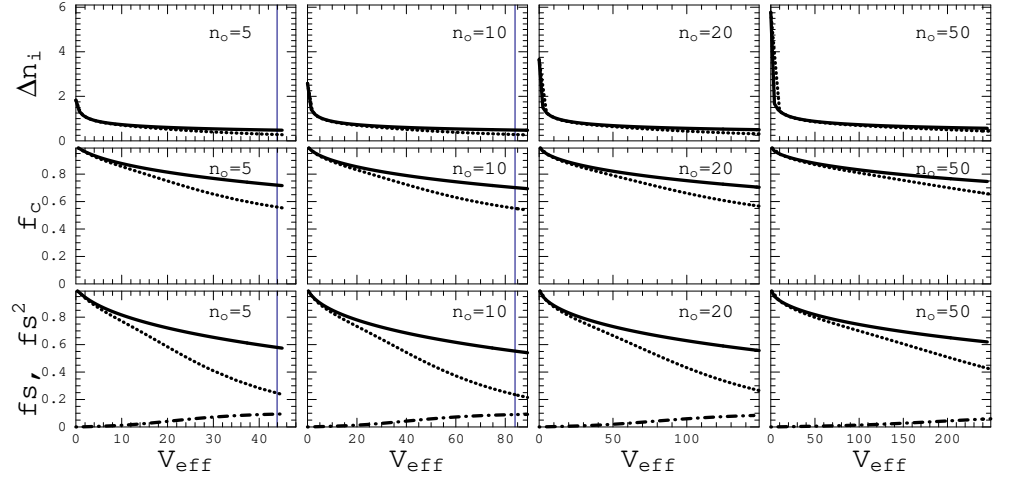}}
\caption{Comparisons of the exact solution (dotted line) and
HFB-Popov (solid line) as a function of $V_{eff}=V/J$, for a
system with $I=3$ and filling factors $n_{o}=5,10,20,$ and $50$.
Top: number fluctuations, middle: condensate fraction, bottom:
superfluid fraction $f_{\mathrm{s}}$. The exact second order term
(dashed line) of the superfluid fraction, $f_{\mathrm{s}}^{(2)}$
is also shown in these plots. The vertical line shown in some
plots is an estimation of $V_{eff}^{crit}.$} \label{Fig1}
\end{figure}

\begin{figure}[tbh]
{\includegraphics[width=7.2 in]{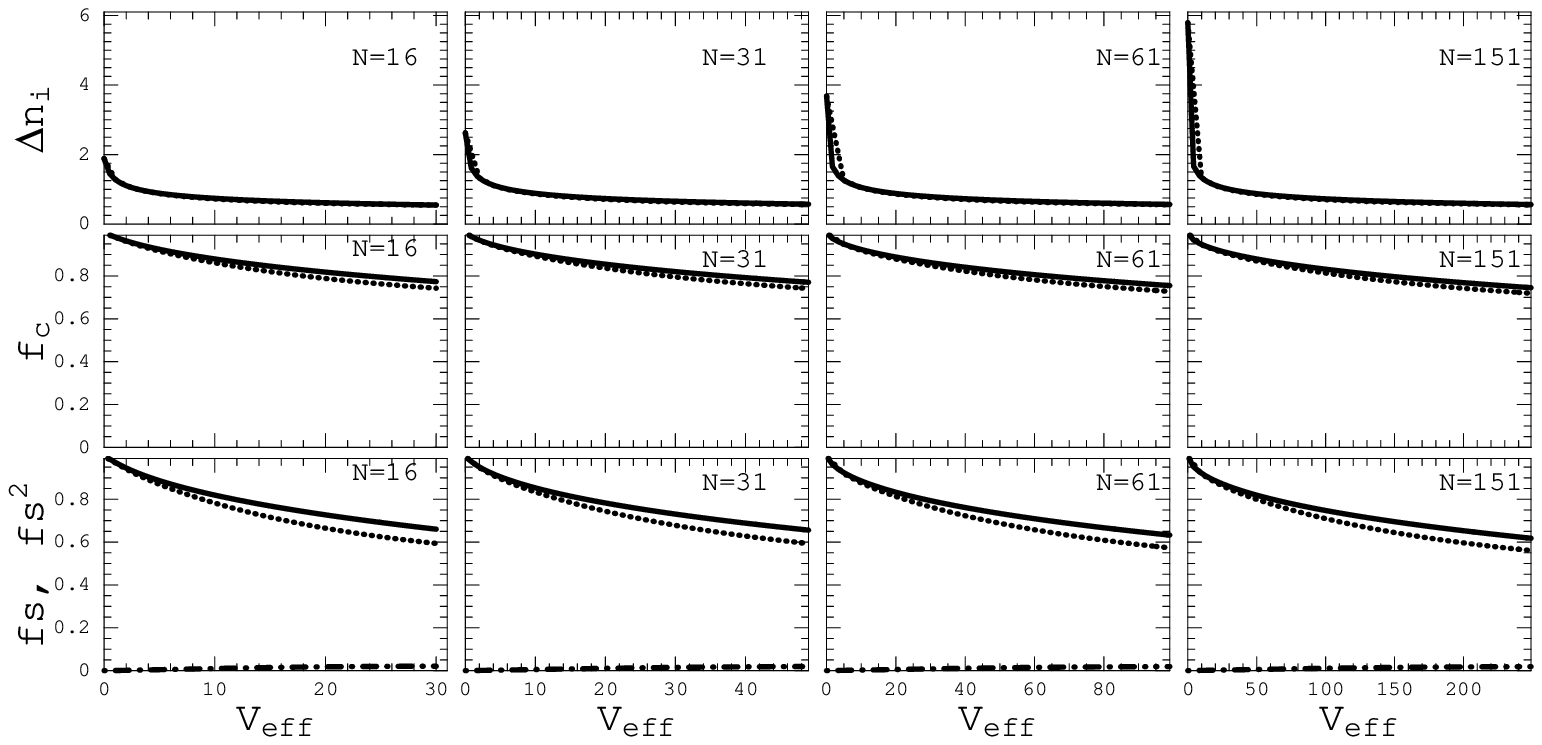}}
\caption{Comparisons of the exact solution (dotted line) and
HFB-Popov (solid line)
for a system with $I=3$ and non commensurate filling factors $%
N=16,31,61,151$ \ as a function of $V_{eff}=V/J$. Top: number
fluctuations, middle: condensate fraction, bottom: superfluid
fraction $f_{\mathrm{s}}$. In these plots the exact second order
term of the superfluid fraction is also shown with a dashed line.}
\label{Fig1/2}
\end{figure}

In Fig. \ref{Fig1}\ we compare the \ number \ fluctuations on a lattice
site, the condensate fraction and the total and second order superfluid
fraction determined from the exact solution of the Bose-Hubbard Hamiltonian
to the self consistent HFB-Popov predictions as a function of the ratio $%
V_{eff}=V/J$. The systems used for the comparisons have three
wells, $I=3$, and commensurate filling factors $n_{0} = 5,10,20$
and 50. Similar results for
the incommensurate case with $N=16,31,61,151$ are shown in Fig. \ref{Fig1/2}%
. We were restricted to consider only three wells due to computational
limitations. The size of the matrix needed in the exact solution for $N$
atoms and $I$ wells scales as $\frac{(N+I-1)!}{N!(I-1)!}$. However, if the
HFB-Popov approach works well for these small systems we expect it to
provide a good description of the larger systems prepared in the lab.

Because the second order term \ of the superfluid fraction (second
term of Eq.\ref{sffraction}) vanishes in the HFB-Popov approach
(see \ Eq. \ref {HFBsf}), we only expect the self consistent
HFB-Popov theory to give a good description of the superfluid
fraction in the region where the second order term is extremely
small, provided it predicts accurately the first order term. This
is exactly what is observed in the plots. When the second order
term starts to grow, typically above
$0.5V_{eff}^{{}_{\text{crit}}}$, the HFB-Popov theory starts to
fail. An estimate of $V_{eff}^{\text{crit}}$ is shown by a
vertical line in some of the figures. This was obtained by using
the second order perturbative approach presented in Ref.
\cite{VanOosten}. With increasing filling factor the critical
value is shifted towards larger values of the interaction
strength, and the region in which the HFB-Popov theory is accurate
gets larger. It is interesting to note that the number
fluctuations predicted by the theory are accurate in a greater
range than the other physical quantities shown. Its predictions of
squeezing agree very well with the exact solutions \ right up to
the point\ where the number fluctuations become less than unity.

For the cases with non-commensurate fillings depicted in Fig. \ref{Fig1/2},
the agreement is significantly better for all quantities. This is not
surprising because when the filling is not commensurate there is always a
superfluid present and the Mott transition doesn't occur. As can be seen in
the plots for these cases the second order term is always very small.

\subsection{Inhomogeneous lattice}

In this section we consider the experimentally relevant case when there is
an external magnetic confinement in addition to the lattice potential. In
this situation, the self consistent HFB-Popov equations take the form:
\begin{eqnarray}
{\small \hbar \omega }_{q}{\small u}_{i}^{q}{\small +c}^{q}{\small z}_{i} &=&%
{\small (2V(|z}_{i}{\small |}^{2}{\small +\tilde{n}}_{i}{\small )-\mu
+\Omega i}^{2}{\small )u}_{i}^{q}{\small -J(u}_{i+1}^{q}{\small +u}_{i-1}^{q}%
{\small )-Vz}_{i}^{2}{\small v}_{i}^{q}{\small ,}  \label{equt} \\
{\small -\hbar \omega }_{q}{\small v}_{i}^{q}-{\small c}^{q}{\small z}%
_{i}^{\ast } &=&{\small (2V(|z}_{i}{\small |}^{2}{\small +\tilde{n}}_{i}%
{\small )-\mu +\Omega i}^{2}{\small )v}_{i}^{q}{\small -J(v}_{i+1}^{q}%
{\small +v}_{i-1}^{q}{\small )-Vz}_{i}^{\ast 2}{\small u}_{i}^{q}, \\
{\small \mu z}_{i} &=&{\small -J(z}_{i+1}{\small +z}_{i-1}{\small )+}\left(
V(|z_{i}|^{2}+2\tilde{n}_{i})+\Omega i^{2}\right) {\small z}_{i}{\small ,}
\label{eqz2} \\
\tilde{n}_{i} &=&\sum_{q}|v_{i}^{q}|^{2}, \\
N &=&\sum_{i}(|z_{i}|^{2}+\tilde{n}_{i}).  \label{eqconstr} \\
c^{q} &=&V\sum_{i}|z_{i}|^{2}(z_{i}^{\ast }u_{i}^{q}-z_{i}v_{i}^{q})
\end{eqnarray}
where $\Omega =\frac{1}{2}m\omega ^{2}a^{2}$, with $m$ the mass of
the atoms, $\omega $ the trap frequency, and $a$ the lattice
spacing. The site indices $i$ are chosen such that $i=0$
corresponds to the center of the trap. Again the $c^{q}\ $are
introduced to ensure the orthogonality of the excitations to the
condensate \cite{Morgan}. We have solved the HFB-Popov equations
for this system by an iterative procedure, similar to the one
followed in Ref. \cite{Dodd}. Each cycle of the iteration consists
of two steps. In the first step we solve Eq. (\ref{eqz2}) subject
to the constraint Eq. (\ref{eqconstr}) by using the
${\tilde{n}_{i}}$ obtained in the previous cycle. This generates
new values for the ${z_{i}}$. In the
second step we solve for $\left\{ u_{i}^{q},v_{i}^{q}\right\} $ in Eqs. (%
\ref{equt}) using the ${\tilde{n}_{i}}$ from the previous cycle
and the newly generated ${z_{i}}$. The $\left\{
u_{i}^{q},v_{i}^{q}\right\} $ are used then to update
${\tilde{n}_{i}}$. Because the HFB-Popov is gapless, it is
possible to keep the orthogonality of the excitations to the
condensate by solving Eqs. (\ref{equt}) with the $c^{q}$ set to
zero\ but removing in each cycle the projection of the calculated
$\left\{ u_{i}^{q},v_{i}^{q}\right\} $ amplitudes onto the
condensate. Convergence is reached when the change in
$\sum_{i}|\tilde{n}_{i}|^{2}$ from one cycle to the next is
smaller than a specified tolerance.

The parameters chosen for the numerical calculations were $\Omega
=0.0015E_{R}$, with $E_{R}$ the one photon recoil energy, which
for the case of a Rubidium condensate corresponds to a trap
frequency of approximately 90 Hz. \ We used a total number of 1000
atoms, $N=1000,$ and set $VN=1.0E_{R}$. $J$ \ was varied to
achieve a range of $\ V_{eff}=V/J$ \ between 0.01 and 312. The
range was chosen based on a local mean field approach \cite
{VanOosten}, which for our parameters \ estimates the transition
region between $V_{eff}\approx640$ (at the center where the local
filling factor is approximately 80) and $V_{eff}\approx12$ (at the
wings).

\begin{figure}[tbh]
\includegraphics[width=6.2in]{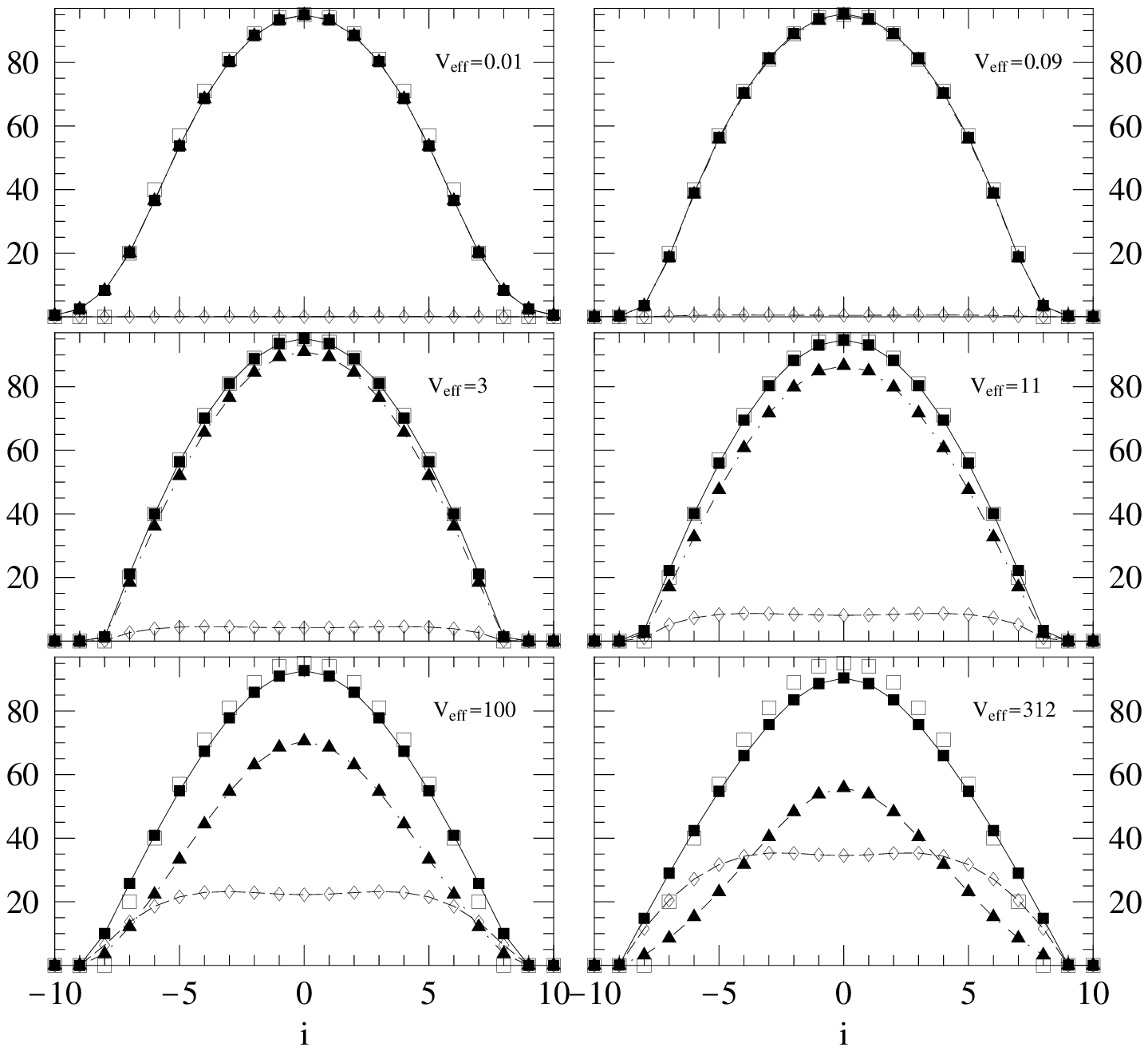}
\caption{Condensate density (triangles), total \ density (filled
boxes) \ and local depletion (empty diamonds) as a function of the
lattice site for different values of $V_{eff.}$ Although these
quantities are defined only at the discrete lattice sites \ we
join them \ to help visualization. The empty boxes represent the
exact solution for the case J=0.} \label{Fig2}
\end{figure}

\begin{figure}[tbh]
{\centering \includegraphics[width=5.2 in]{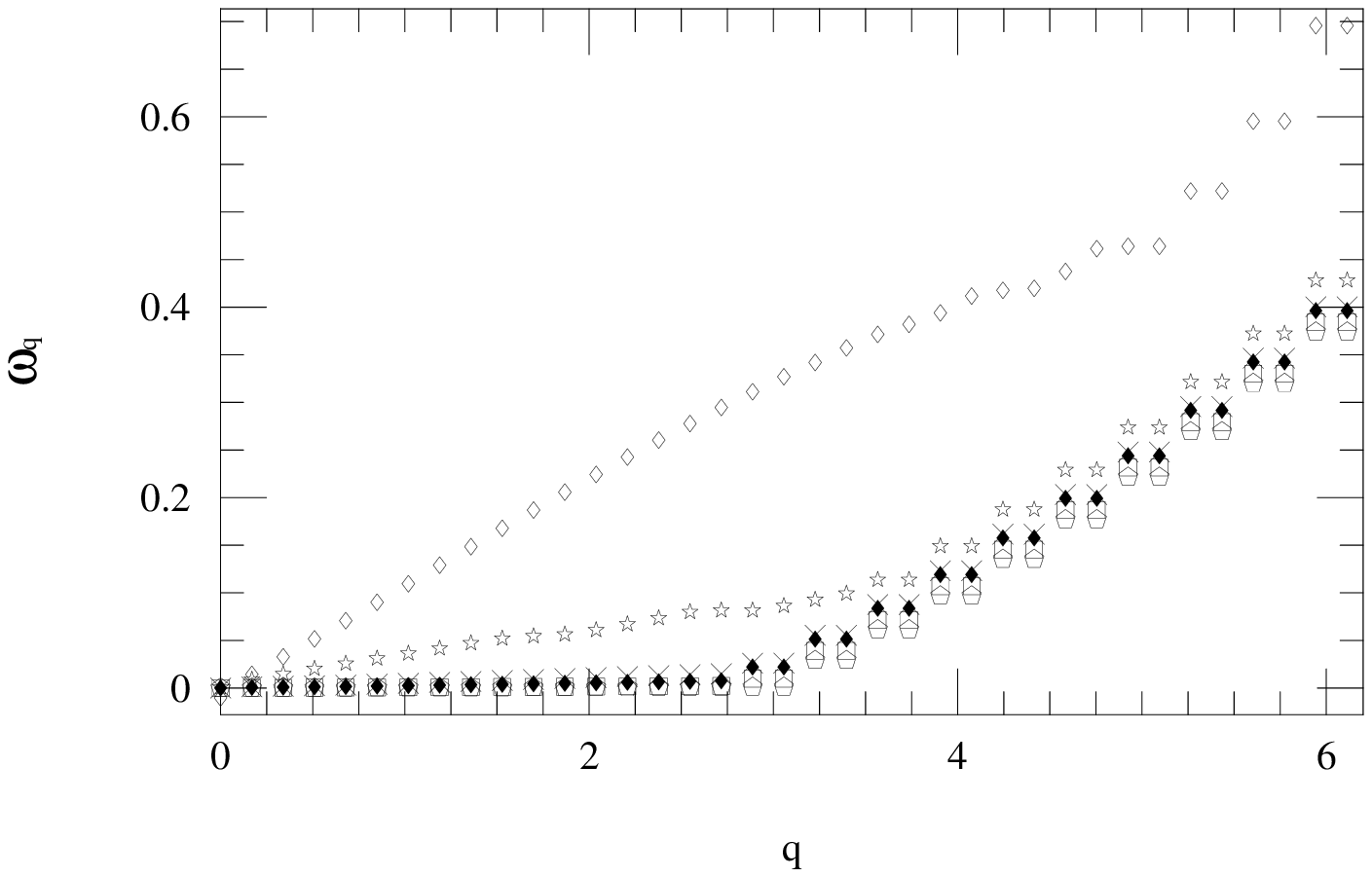} }
\caption{Quasiparticle spectrum in ascending order predicted by
the HFB-Popov theory for
different values of $V_{eff}$: Empty diamonds $(V_{eff}=0.01)$, stars ($%
V_{eff}=0.09$), crosses $(V_{eff}=3)$, filled diamonds
$(V_{eff}=11)$, empty boxes $(V_{eff}=100)$ and polygons
$(V_{eff}=312)$ } \label{Fig3}
\end{figure}

\begin{figure}[tbh]
{\centering \includegraphics[width=5.2in]{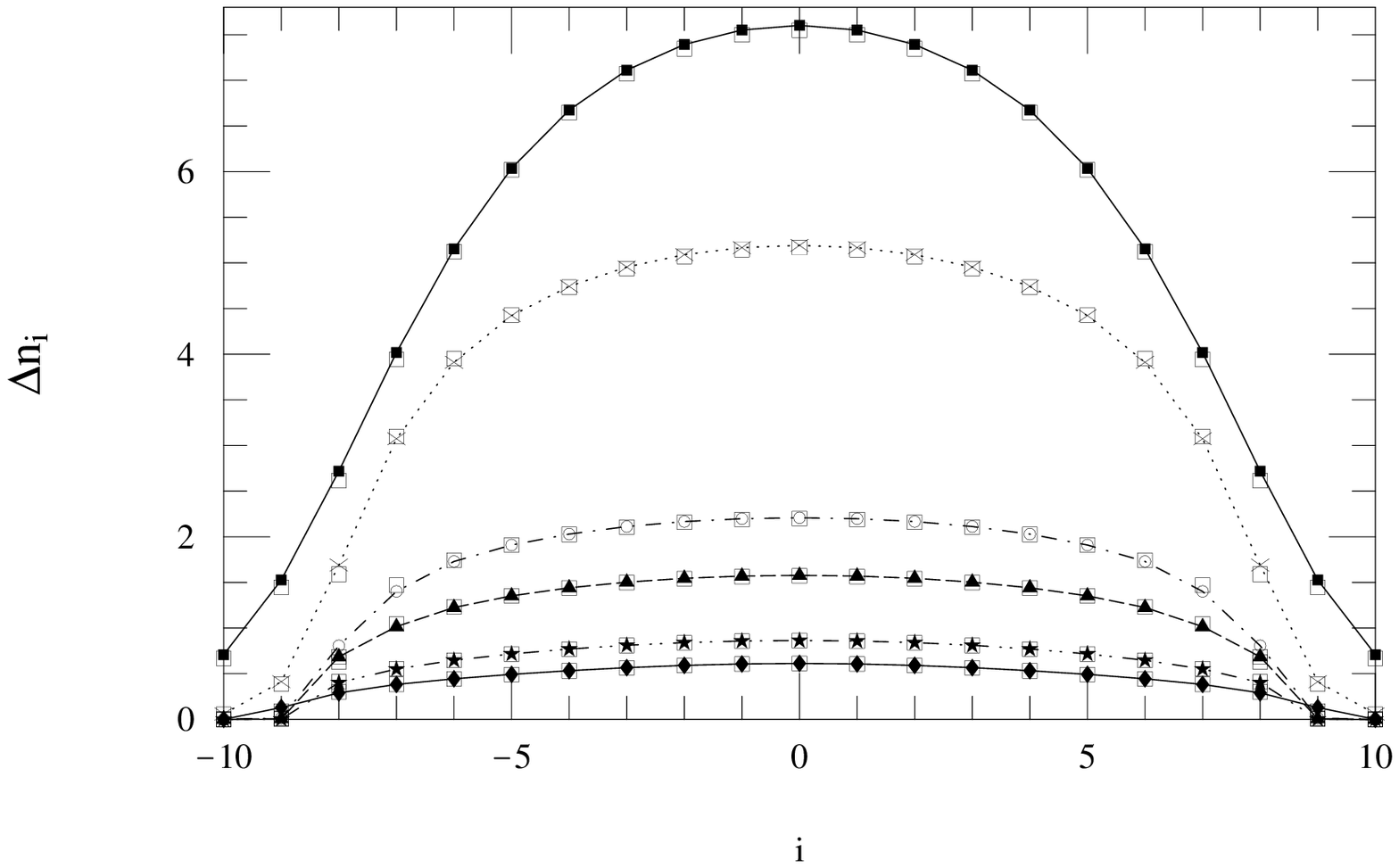} }
\caption{Number fluctuations in the self consistent HFB-Popov
approach as a function of lattice site for $V_{eff} = 0.01$
(boxes), $V_{eff} =0.09$ (crosses), $V_{eff} =3$
(circles),$V_{eff} =11$ (triangles), $V_{eff} =100$ (stars) and
$V_{eff} =312$ (diamonds). The \ maximum value reached by the profile decreases as $%
V_{eff}$ \ is increased. The empty boxes shown for each of the
curves correspond to the number fluctuations predicted by the
homogeneous HFB-Popov model using a local density approximation. }
\label{Fig4}
\end{figure}

\bigskip The results of the numerical calculations are summarized in Figs.
\ref{Fig2} to \ref{Fig7}. In Fig. \ref{Fig2} we plot the evolution
of the density profile (black boxes), the condensate population
(triangles) and the on-site depletion (empty diamonds) as $\
V_{eff\text{ }}$ is increased. In the plots we
also show, for comparison purposes, the ground state density profile for $%
J=0 $ (empty boxes). This has the advantage that can be calculated exactly
from the Hamiltonian. In general we observe the reduction of the condensate
population \ and thus the increment of the depletion as the interaction
strength is increased. When the system is in the superfluid regime most of
the atoms are in the condensate but as $J$ is decreased the depletion of the
condensate becomes very important.

For the chosen parameters, the density profile has a parabolic
shape reflecting the confining potential. By comparing the
evolution of the density as $J$ is decreased with the exact
solution at $J=0$, we can crudely estimate the validity of the
HFB-Popov calculations. The density evolves from a Gaussian type
(see plots for $V_{eff\text{ }}=0.01$ and $0.09$) with smooth
edges towards a Thomas-Fermi profile with sharp edges adjusting
its shape to the $J=0$ profile. \ We can appreciate that around
$V_{eff\text{ } }=3$ both profiles are almost equal. For lower
values of $J$ the HFB-Popov density starts to differ from the
$J=0$ one, \ even though the system is closer to the $J=0$ limit.
We can say that beyond this point higher order correlations,
neglected by the theory, begin to be important. The departure of
the HFB-Popov density profile  from the $J=0$ one  as J is
decreased begins at the edges (see panel corresponding to
$V_{eff\text{}}=11$ and $100$). \ This is something expected if we
look at the on-site depletion. For such values of $V_{eff}$ the
local depletion at the wings corresponds to a considerable
percentage of the condensate populations, and thus the validity of
the HFB-Popov assumptions starts to be dubious. The homogeneous
results shown in the previous section corroborate our present
statements for the confined system. For the smallest filling
factor (see Fig.\ref {Fig1}) the differences between the
homogeneous HFB-Popov calculations and the exact solutions become
important for values of \ $V_{eff\text{ }}$ greater than 20.
 For higher values of $V_{eff}$, see plot for $V_{eff\text{ }}=312$, the HFB-Popov
density predictions differs from the $J=0$ solution even at the
central wells. At this point the failure of the method is clear
and a fully quantal method is required.

\bigskip The HFB-Popov quasiparticle spectrum is shown in Fig.\ref{Fig3}. It
can be observed how the lower energy eigenvalues evolve from a
linear non degenerated spectrum to an almost degenerated one as J
is decreased. It is worth it to mention that the small energy
difference between the ground and first excited states for \ high
values of \ $V_{eff\text{ }}$ makes the numerical solution very
unstable in the sense that it was very easy  to jump to an excited
state when solving for the condensate wave function. The decrement
in the energy spacing predicted by the HFB-Popov theory \ as the
system approaches the transition is very useful to keep in mind
for the experimental realization of the Mott transition. As the
optical lattice depth is ramped up the adiabaticity criteria is
harder to fulfill.

In Fig. \ref{Fig4} we plot the results for the number fluctuations
found numerically using the inhomogeneous HFB-Popov approach. The
number fluctuations profile reflects the condensate profile. We
also show  the number fluctuations evaluated   by using a local
density approximation (empty boxes). The latter was calculated by
substituting in the number
fluctuations expression  (Eq. \ref {numflu}) the  $%
\{u^{q},v^{q}\}$ amplitudes  found for the  homogeneous system
(Eqs. \ref{ushfb} and \ref{eqv}), but replacing the condensate
density in each lattice site by the one found numerically for the
trapped system (see Fig. \ref{Fig2}). The complete agreement
between the two approaches justifies the validity of the local
density approximation for the estimations of local quantities in
confined systems. Based on this agreement and the results for the
homogeneous system shown in the previous section, we expect that
the inhomogeneous  HFB-Popov results for squeezing also agrees
with the exact solution right up to the transition.
\bigskip

\subsection{Quasi-momentum distribution in the inhomogeneous lattice}

\begin{figure}[tbh]
{\centering \includegraphics[width=6.2 in]{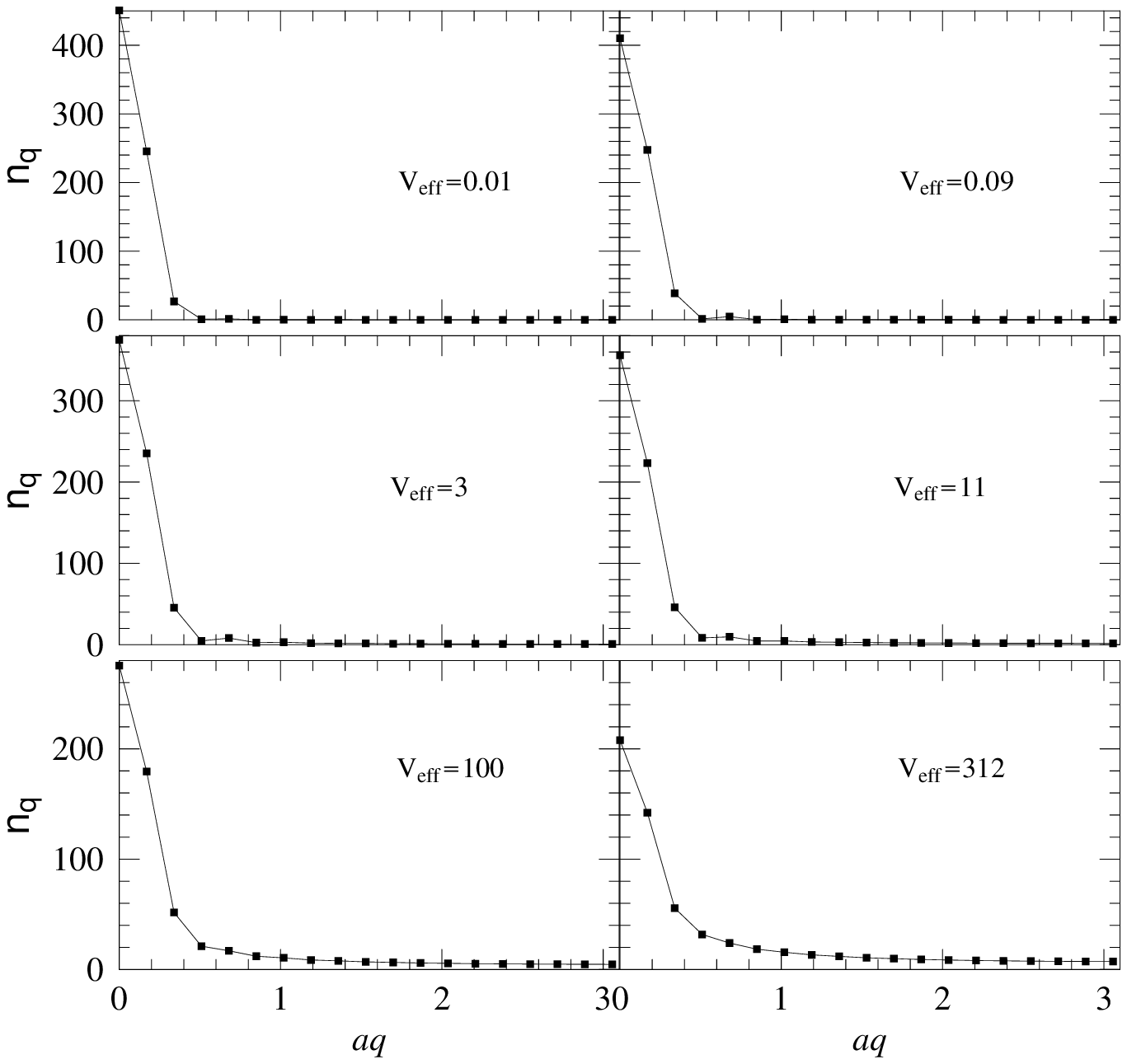} }
\caption{Quasi-momentum distribution as a function of $q a$, $a$
the lattice spacing, $q$ the quasimomentum, for different values
of $V_{eff.}$} \label{Fig5}
\end{figure}

The quasi-momentum distribution of the atoms released from the
lattice is important because it is one of the most easily
accessible quantities to the experiments. The quasi-momentum
distribution function $n_{q}$ is defined as \cite{RBSuperfluid}
\begin{equation}
n_{q}=\sum_{i,j}e^{\mathrm{i}q(i-j)a}\langle a_{i}^{\dagger }a_{j}\rangle ,
\end{equation}
where the quasi-momentum $q$ can assume discrete values which are integer
multiples of $\frac{2\pi }{Ia}$, $a$ is the lattice spacing. In Fig. \ref
{Fig5} we present the quasi-momentum distribution for the same parameters
used in the previous section. The distribution for the two lowest values of $%
V_{eff}$ corresponds to the one that characterizes an uncorrelated
superfluid phase with a narrow peak at small quasi-momenta. As the
hopping rate is decreased we observe that the sharpness of the
central peak decreases and the distribution extends towards large
quasi-momenta. It is interesting to note the appearance of a small
peak between $q=0.5$ and $1$ which is most noticeable for the
$V_{eff}=3$ case. This agrees with the results found in
\cite{Kashurnikov} where they solve numerically the Bose-Hubbard
Hamiltonian by using Monte Carlo simulations. We attribute the
origin of the small peak to the depletion of the condensate at the
wings. For the parameters when the small peak is present, the most
important contribution to the quasi-momentum distribution still
comes from the condensate atoms. The step function like shape of
the condensate profile causes an oscillatory \ $|\sin (x)/x|$
shape of the quasi-momentum distribution. As the lattice depth is
increased the hopping becomes energetically costly, the long-range
order starts to decrease and the Fourier spectrum becomes broader.

\begin{figure}[tbh]
{\centering \includegraphics[width=6.2 in]{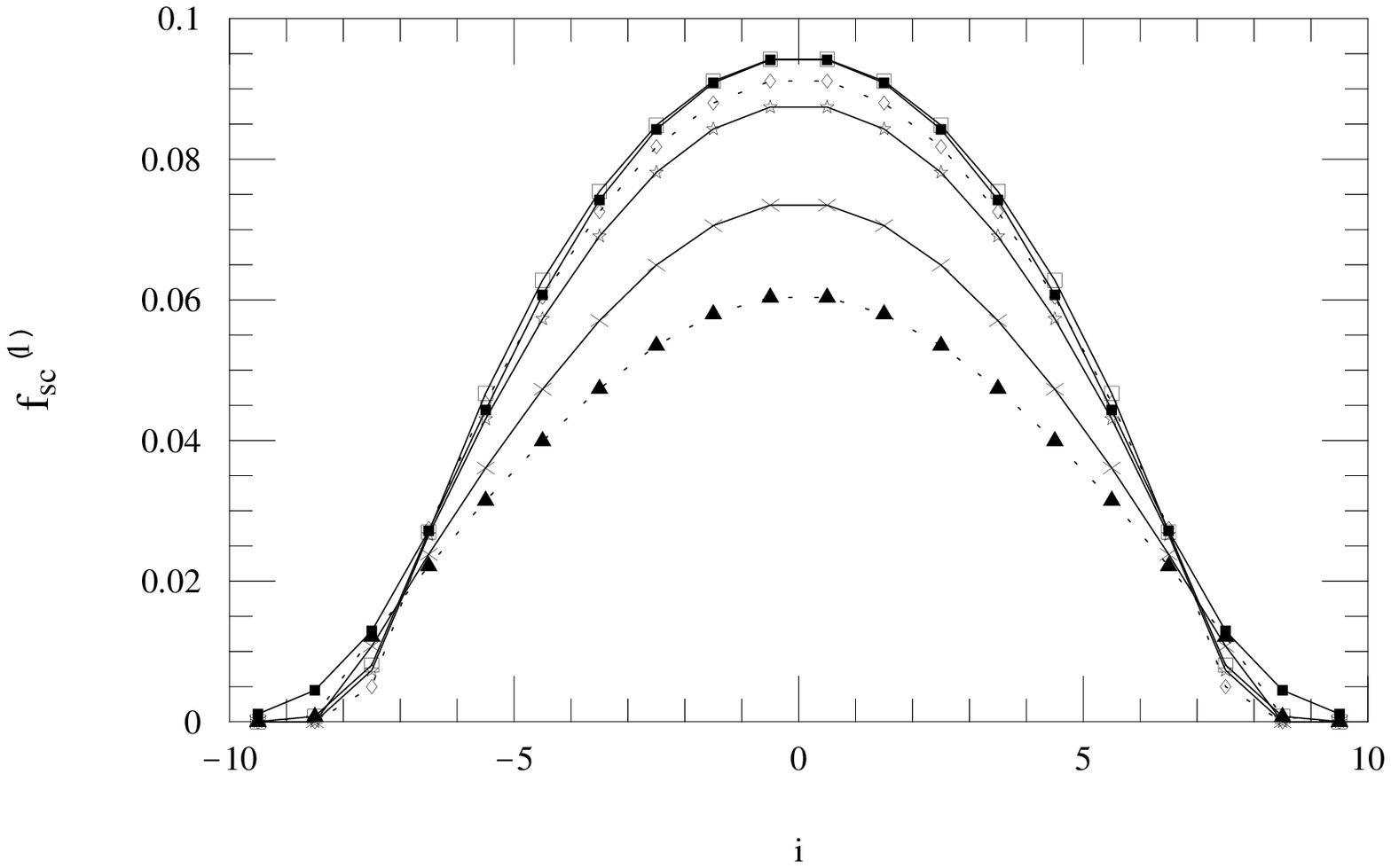} }
\caption{First order on-site superfluid fraction as a function of
the lattice site for different values of $V_{eff\text{.}}$ Filled
boxes: $V_{eff\text{.}}$ =0.01, empty boxes:
$V_{eff\text{.}}=0.09$, empty diamonds: $V_{eff\text{.}} =3$,
stars: $V_{eff\text{.}}=11$, crosses: $V_{eff\text{.}}=100$ and
triangles: $V_{eff\text{.}}=312$.} \label{Fig6}
\end{figure}

\bigskip
\begin{figure}[tbh]
{\centering \includegraphics[width=5.2 in]{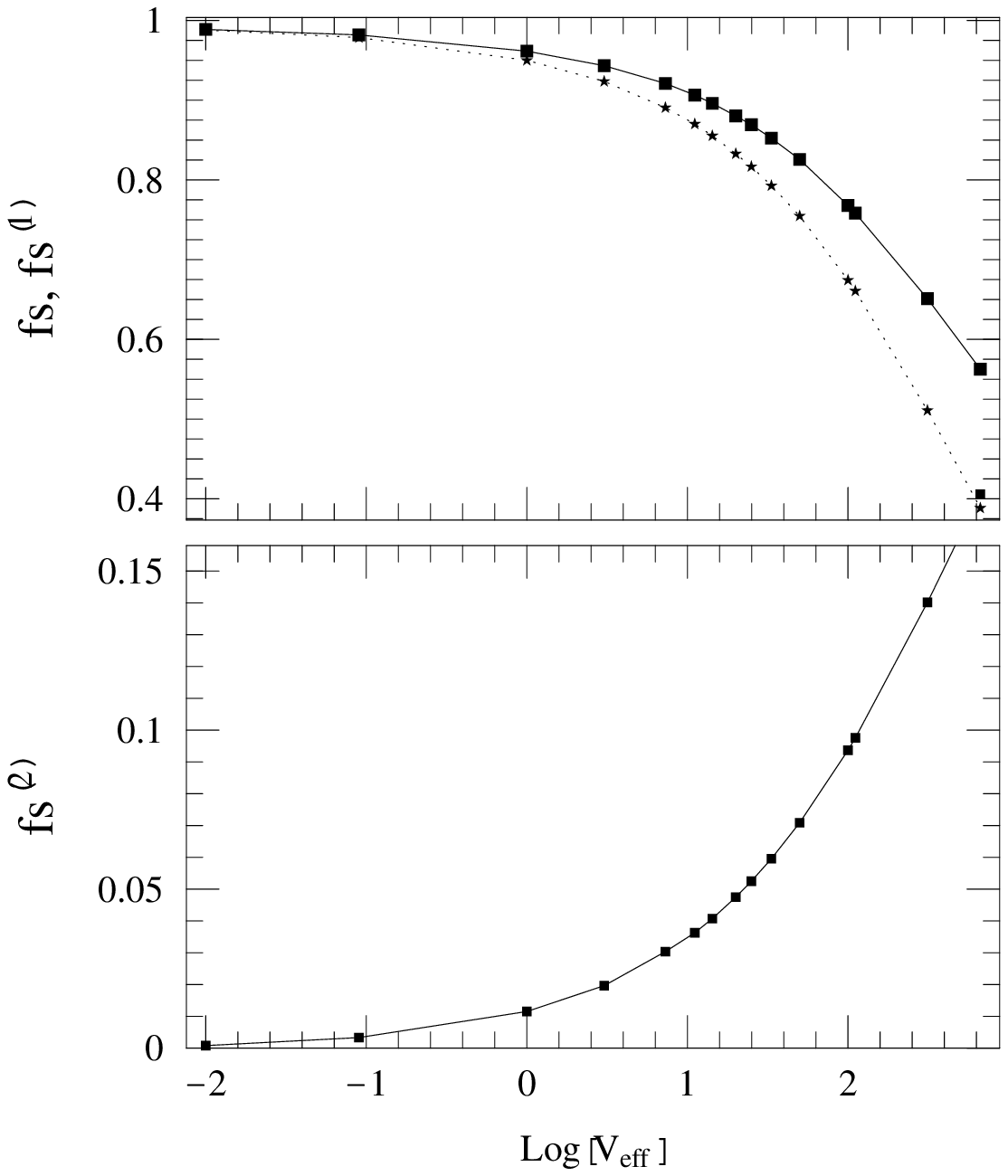} }
\caption{Top panel: First order (boxes) and total (stars)
superfluid fraction as a function of $V_{eff.\text{ }}$. Bottom
panel : second order superfluid fraction as a function of
$V_{eff.\text{ }}$All these quantities are calculated in the self
consistent HFB- Popov approach.} \label{Fig7}
\end{figure}

In Fig. \ref{Fig6} we plot the first order on site superfluid
fraction $f_{si}^{(1)}$ which was defined in Eq.(\ref {HFBsf}).
The curves corresponding to $V_{eff}=0.01-11$, which are in the
regime where the HFB-Popov is expected to be valid, \ depict \ how
as \ $V_{eff}$ \ is increased the superfluid  profile decreases
faster at the wings and at the center but no major change is
observed in the middle section. The evolution of the on-site
superfluidity as  the interaction strength is increased,
exhibiting a domain localized decrement instead of a global one,
is in agreement with the development of \ uncompressible regions
surrounded by superfluid rings predicted for trapped systems
\cite{Batrouni}  as the transition is approached.

Fig. \ref{Fig7} shows the first order and total superfluid
fraction and also the second order superfluid fraction as a
function of $V_{eff}$. Different from the translationally
invariant case, the second order contribution calculated in the
HFB-Popov approach doesn't vanish for the inhomogeneous
system. The rapid decrement of the superfluid fraction observed \ after $%
Log(V_{eff})\sim 1.2$ \ is a signature that the system is entering
a highly correlated regime. Beyond this point higher order
correlations neglected in the HFB-Popov approach  become crucial
and a more accurate approach is required.

The Mott transition is a quantum phase transition and as all
critical phenomena its behavior depends strongly on the
dimensionality of the system. In the present analysis, due to
computational limitations, we considered  one dimensional systems.
Experimentally, the Mott transition has been achieved
\cite{BlochMunich} in a 3 dimensional lattice\ with filling
factors between 1 and 3. Even though the HFB-Popov approach fails
to describe the strong coupling regime  for the one dimensional
systems we considered in the present paper, we showed how the
method is incredibly powerful in describing most of its
characteristic features  as they are driven from the superfluid
regime towards the transition. We expect the HFB-Popov method to
give a better description of the transition as the dimensionality
of the system is increased and therefore to be a good model  in an
experimental situation.

 As shown in previous studies \cite
{Fisher}, \cite{G. Batrouni} the Mott transition in\ a
d-dimensional homogeneous system has two different critical
behaviors: one (d+1) XY- like, for systems with\ fixed\ integer
density as the interaction strength is changed, and one\ mean
field-like exhibited when the transition is induced by changing
the density. Different from the homogeneous case where \ the Mott
transition is characterized by the global offset of the
superfluidity, for confined systems, commensuration is only well
defined locally. The inhomogeneity introduced by the confined
potential \ allows the existence of extended Mott domains  (above
a critical interaction strength) surrounded by superfluid ones
\cite{Batrouni}, thus the total superfluid fraction doesn't
vanishes in the Mott regime. This issue, together with the fact
that the finite length scale introduced by the trap suppresses the
long wave fluctuations which are responsible for destroying the
mean
field \cite{Morgan}\footnote{%
One obvious consequence of this is that BEC is possible in one and
two dimensions in a trap whereas in the homogeneous, thermodynamic
limit it can not occur in fewer than three dimensions}, make us
believe   the critical behavior in confined systems  to be more
mean-field like. Because the critical dimension for the latter
type of \ transition is two \cite {Fisher}, \cite{G. Batrouni}, we
expect that for trapped systems in $d=3$, the range of validity of
the HFB-Popov extends closer to the transition.

\section{Summary}

We have developed in this article a Bogoliubov method for describing the
approach of a condensate loaded in an optical lattice towards the Mott
transition. We have shown that this method can be used to predict the
relevant physical quantities over a useful range. We have also shown how it
gives a powerful insight into the way quantum depletion reduces the long
range order and the superfluid fraction.

\section*{Acknowledgments}

This work was supported in part by US National Science Foundation grants
PHY-0100634 and PHY-0100767, the United Kingdom's Engineering and Physical
Sciences Research Council, the Cold Quantum Gases Network Research Training
Network, and the Advanced Research and Development activity.

%%%%%%%%%%%%%%%%%%%%%%%%%%%%%%%%%%%%%%%%%%%%%%%%%%%%%%%%%%%%%%%%%%%%%%%%

\end{document}